\documentclass[conference, letterpaper]{IEEEtran}
\IEEEoverridecommandlockouts
\usepackage{amsmath,amssymb,amsfonts}
\usepackage{graphicx}
\usepackage{textcomp}
\usepackage{xcolor}
\def\BibTeX{{\rm B\kern-.05em{\sc i\kern-.025em b}\kern-.08em T\kern-.1667em\lower.7ex\hbox{E}\kern-.125emX}}
\usepackage{svg}
\usepackage[T1]{fontenc} 
\usepackage{microtype} 

\usepackage[english]{babel} 
\usepackage{microtype} 

\usepackage[english]{babel} 

\usepackage{caption}
\usepackage{subcaption}
\usepackage{siunitx}
\usepackage{gensymb}
\usepackage{booktabs}
\usepackage{adjustbox}
\usepackage{tikz}
\usepackage{svg}
\usepackage{soul}
\usepackage{tabularx}
\usepackage{algorithm}
\usepackage{algpseudocode}
\usepackage{algorithmicx}
\graphicspath{ {images/} }
\usepackage[font=small,labelfont=bf]{caption}

\usepackage{csquotes}
\usepackage{glossaries}
\usepackage{subcaption}
\glsdisablehyper
\newacronym{rl}{RL}{Reinfocement Learning}
\newacronym{mdp}{MDP}{Markov Decision Process}
\newacronym{sac}{SAC}{Soft Actor-Critic}
\newacronym{ml}{ML}{Machine Learning}
\newacronym{ai}{AI}{Artificial Intelligence}
\newacronym{rl3}{RL-Zoo3}{RL Baselines3 Zoo}
\newacronym{sgd}{SGD}{Stochastic Gradient Descent}
\newacronym{marl}{MARL}{Multi-Agent Reinforcement Learning}
\newacronym{dl}{DL}{Deep Learning}
\newacronym{nn}{NN}{Neural Network}
\newacronym{gnn}{GNN}{Graph Neural Network}
\newacronym{gat}{GAT}{Graph Attention Network}
\newacronym{diga}{DiGA-DQN}{Dueling Graph Attention Deep Q-Network}
\newacronym{olsr}{OLSR}{Optimized Link State Routing}
\newacronym{cnn}{CNN}{Convolutional Neural Network}
\newacronym{sg}{SG}{Stochastic Game}
\newacronym{ctde}{CTDE}{Centralized Training Decentralized Execution}
\newacronym{dtde}{DTDE}{Decentralized Training Decentralized Execution}
\newacronym{gcn}{GCN}{Graph Convolutional Network}
\newacronym{mpr}{MPR}{Multipoint Relay}
\newacronym{ilp}{ILP}{Integer Linear Program}
\newacronym{mcds}{MCDS}{Minimum Connected Dominating Set}
\newacronym{cds}{CDS}{Connected Dominating Set}
\newacronym{posg}{POSG}{Partially Observable Stochastic Game}
\newacronym{dpomdp}{Dec-POMDP}{Decentralized Partially Observable Markov Decision Process}
\newacronym{pomdp}{POMDP}{Partially Observable Markov Decision Process}
\newacronym{manet}{MANET}{Mobile Ad-hoc Network}
\newacronym{vanet}{VANET}{Vehicular Ad-hoc Network}
\newacronym{ddqn}{DDQN}{Double Deep Q-Network}
\newacronym{td}{TD}{Temporal Difference}
\newacronym{mlp}{MLP}{Multi Layer Perceptron}
\newacronym{ldyan}{L-DyAN}{Local Dynamic Attention Network}
\newacronym{hldyan}{HL-DyAN}{Hyperlocal Dynamic Attention Network}
\newacronym{dgn}{DGN}{Graph Convolutional Reinforcement Learning}
\newacronym{tc}{TC}{Topology Control}
\newacronym{smf}{SMF}{Simplified Multicast Forwarding}
\newacronym{uav}{UAV}{Unmanned Aerial Vehicle}
\newacronym{lstm}{LSTM}{Long Short-Term Memory}
\newacronym{lvc}{LVC}{Live Virtual Constructive}
\begin{document}

\title{Distributed Autonomous Swarm Formation for Dynamic Network Bridging}
\author{
\IEEEauthorblockN{
Raffaele Galliera \IEEEauthorrefmark{1}\IEEEauthorrefmark{2},
Thies Möhlenhof \IEEEauthorrefmark{3},
Alessandro Amato \IEEEauthorrefmark{1}\IEEEauthorrefmark{2}, \\
Daniel Duran \IEEEauthorrefmark{1},
Kristen Brent Venable \IEEEauthorrefmark{1}\IEEEauthorrefmark{2},
Niranjan Suri \IEEEauthorrefmark{1}\IEEEauthorrefmark{2}\IEEEauthorrefmark{4}}

\IEEEauthorblockA{\textit{\IEEEauthorrefmark{1}Institute for Human \& Machine Cognition (IHMC)}\\}

\IEEEauthorblockA{
    \textit{\IEEEauthorrefmark{2}Department of Intelligent Systems \& Robotics} -
    \textit{The University of West Florida (UWF)}\\
Pensacola, FL, USA
}

\{rgalliera, aamato, dduran, bvenable, nsuri\}@ihmc.org

\IEEEauthorblockA{\textit{\IEEEauthorrefmark{3}Fraunhofer Institute for Communication, Information Processing and Ergonomics (FKIE)} \\
Wachtberg, Germany \\
thies.moehlenhof@fkie.fraunhofer.de
}

\IEEEauthorblockA{\textit{\IEEEauthorrefmark{4}US Army Research Laboratory (ARL)} \\
Adelphi, MD, USA \\}
}

\maketitle

\begin{abstract}


Effective operation and seamless cooperation of robotic systems are a fundamental component of next-generation technologies and applications. In contexts such as disaster response, swarm operations require coordinated behavior and mobility control to be handled in a distributed manner, with the quality of the agents' actions heavily relying on the communication between them and the underlying network. 
In this paper, we formulate the problem of dynamic network bridging in a novel Decentralized Partially Observable Markov Decision Process (Dec-POMDP), where a swarm of agents cooperates to form a link between two distant moving targets. Furthermore, we propose a Multi-Agent Reinforcement Learning (MARL) approach for the problem based on Graph Convolutional Reinforcement Learning (DGN) which naturally applies to the networked, distributed nature of the task. The proposed method is evaluated in a simulated environment and compared to a centralized heuristic baseline showing promising results. Moreover, a further step in the direction of sim-to-real transfer is presented, by additionally evaluating the proposed approach in a near Live Virtual Constructive (LVC) UAV framework.
\end{abstract}



\section{Introduction}
In the evolving landscape of wireless communication, deploying autonomous agents, particularly \glspl{uav}, presents unique opportunities for establishing dynamic communication networks. Scenarios where traditional infrastructure is unavailable or impractical, such as disaster response or remote area connectivity, are just a few direct application examples. The mobility and flexibility of \glspl{uav} allow for creating an adaptive and resilient network topology, responding to the ever-changing environmental conditions. 

Consider, for example, two mobile, independent nodes engaged in some task. To accomplish their duties, it might be essential for these two nodes to exchange information, even for just a short period. However, the infrastructure needed to ensure their communication link might be unavailable, unreliable, or absent, and, depending on the scenario, centralized solutions governing the \glspl{uav} strategies could be unfeasible or even undesirable. Autonomous \glspl{uav} able to coordinate on distributed strategies might provide a practical solution in ensuring an ad-hoc connection between these two nodes.

This paper introduces a novel \gls{dpomdp} setting to address the challenge of dynamically forming communication links between moving targets using a swarm of \glspl{uav}, which we call \textit{dynamic network bridging}. In this scenario, each \gls{uav}, equipped with limited sensing capabilities, must cooperatively navigate and position itself within the environment to ensure continuous connectivity between the moving targets. Moreover, we propose a decentralized \gls{marl} approach exploiting the networked nature of the task to enable effective cooperation. By employing \gls{dgn}~\cite{Jiang2020Graph} with \glspl{gat}~\cite{veličković2018graph} and \gls{lstm}~\cite{HausknechtS15}, our agents utilize spatio-temporal information while communicating learned latent representations with neighboring agents and target entities to ground their actions. We compare our method with a centralized heuristic, showing how our \gls{marl} approach drives a competitive distributed solution. 
Finally, we present the integration of our \gls{lvc} \gls{uav} framework with our \gls{marl} environment, enabling the deployment of our learned strategies in simulated-only, real-only, or a mixture of real and simulated agents operating in the same environment. 





The remainder of this paper is structured as follows: in Section \ref{sec:rel-work} we discuss related work; Section~\ref{sec:method} provides a detailed description of our \gls{dpomdp} formulation, followed by the description of our \gls{marl} approach in Section~\ref{sec:approach}. We continue with the integration of our \gls{marl} framework with the \gls{lvc} in section \ref{sec:lvc}. Section~\ref{sec:experiments} outlines the experimental setup, including the simulation environment and training procedures. Finally, we conclude with a discussion of future research directions and potential extensions of our approach.


\section{Related Work}
\label{sec:rel-work}
The development of cooperative architectures involving multiple \glspl{uav} has been a significant area of research, particularly in the context of vehicular~\cite{Zhou2015} and tactical~\cite{Tortonesi2012} networks. These studies have primarily focused on addressing communication challenges and distributed decision-making strategies within multi-UAV systems.
The challenge of optimized \gls{uav} placement is explored in ~\cite{Park2016DroneFA}, where the problem is reduced to a packing problem. The authors designed an iterative algorithm to optimize area coverage, network capacity, and routing constraints while minimizing overlapping areas. As an alternative to combinatorial optimization approaches, recent research has shifted towards model-free, learning-based strategies for \gls{uav}-cells management. For instance, Hammami et al.~\cite{Hammami2019} introduce a semi-centralized, multi-agent framework. In this model, a central entity gathers joint actions, while the environmental state encompasses network parameters, \gls{uav} statuses, battery levels, and bandwidth capacities.

Another innovative approach is presented in~\cite{Jiang2021}, which proposes a \gls{marl} approach for \gls{uav}-assisted Roadside Units (RSUs) in providing \glspl{vanet} along highways. In their problem formulation, \glspl{uav} are placed in a straight line at a fixed altitude and can move backward and forward with the primary object of maximizing the number of sub-segments that satisfy a delay constraint. The authors employ independent Deep Q-Learning (DQN)~\cite{marl-book} with shared policies parameterized by small neural networks to periodically update the \glspl{uav} positions. Similar to our approach, the observation of each agent is augmented with information from its immediate neighbors. However, instead of concatenating the neighboring agents' observations, we propose a more flexible approach employing \glspl{gat} and their inherent message-passing mechanism~\cite{gnn}.

Furthermore, we utilize a more cooperative algorithm such as \gls{dgn}, which aims to decompose the agents' value function into agent-wise value functions.


\section{Method}
\label{sec:method}
In this section, we present a \gls{dpomdp} formulation to model the cooperative task of bridging the connection between two moving targets relying on a swarm of $N$ agents in a 2D plane, while perceiving only their local one-hop observation. However, it is worth mentioning that our approach can be expanded to 3D scenarios by augmenting the action space of the agents and refining the attributes of the nodes accordingly.

Let us consider a scenario where each node represents a Mobile Ad-Hoc Networking radio (MANET). Nodes have a specific communication range, representing a certain distance or proximity within which information can be sensed by other nodes. The underlying network can be represented as a dynamic graph $\mathcal{G}(t) = (\mathcal{V}, \mathcal{E}(t))$, where each node represents a mobile node and an edge between two nodes at time $t$ represents the two corresponding nodes being within each other's communication range $r$ at that time. More formally, the set of edges $\mathcal{E}(t)=\{(u,v)|u,v \in \mathcal{V}, distance(u,v,t)\leq r\}$, where \(\emph{{distance}}(u, v, t)\) is a distance metric that measures the spatial or logical distance between nodes \(u\) and \(v\) at time-step $t$.
Hence, for every node \(v \in \mathcal{V}\), the set of its neighbors at time $t$ is defined as $\mathcal{N}_v(t) = \{u \in \mathcal{V}| (v, u) \in \mathcal{E}(t)\}$.


\subsection{A MARL environment for Dynamic Network Bridging}

For multi-agent systems, the \gls{rl} paradigm extends to \gls{marl}~\cite{Busoniu2010}, where multiple entities, potentially learners and non-learners, act within a shared environment. In this context the generalization of \glspl{pomdp} leads to \gls{dpomdp}, characterized by the tuple:
\begin{equation}
\langle \mathcal{I}, \mathcal{S}, {\mathcal{A}^i_{i \in I}, \mathcal{P}, {\mathcal{R}}, \mathcal{O}^i_{i \in I}, \gamma} \rangle
\end{equation}
Here, $\mathcal{I}$ represents the set of agents, $\mathcal{S}$ denotes the state space, ${\mathcal{A}^i}_{i \in \mathcal{I}}$ stands for the action space for each agent, $\mathcal{P}$ is the joint probability distribution governing the environment dynamics given the current state and joint actions, ${\mathcal{R}}$ denotes the reward function, and $\mathcal{O}^i_{i \in \mathcal{I}}$ represents the set of observations for each agent. Such game-theoretic settings are used to model fully cooperative tasks where all agents have the same reward function and share a common reward.
 
In this work, we construct a \gls{dpomdp} formulation for the task of connecting two moving targets relying on a swarm of $N$ agents. Given the network represented by graph \(\mathcal{G}_0 = (\mathcal{V}, \mathcal{E}_0)\) at time $t_0$, and node $T_1, T_2 \in \mathcal{V}$, we define the \gls{dpomdp} associated to the optimized connection of $T_1$ and $T_2$ and the moving target update function $\mathcal{U}_T$, with the tuple:
\begin{equation}
\langle \mathcal{I}, \mathcal{T}, \mathcal{S}, \mathcal{A}^i_{i \in I}, \mathcal{U}_T, \mathcal{P}, \mathcal{R}, \mathcal{O}^i_{i \in I}, \gamma \rangle
\end{equation}
The tuple consists of the following enumerated elements:

\subsubsection{Agent $\mathcal{I}$ and Target $\mathcal{T}$ sets}
$\mathcal{I}$ represents the set of learning agents, and $\mathcal{T}$ denotes the set of moving targets, specifically $T_1$ and $T_2$.


\subsubsection{Observation $\mathcal{O}^i_{i \in \mathcal{I}}$ and State set $\mathcal{S}$}

Each agent $i$ in $\mathcal{I}$ has a local observation $\mathcal{O}^i$, which includes the structure of its neighborhood as well as the features representing its neighbors, as described in Table~\ref{tab:state}. These features include node ID, the current coordinates of that node, the action taken, and the coordinates of the targets that the node aims to connect. The global state $\mathcal{S}$ encompasses the entire graph structure of the network, along with the features describing all agents and targets. However, these are only partially observable by each agent due to their limited observation range.


\subsubsection{Moving Target Update Function $\mathcal{U}_T$}

The moving target update function $\mathcal{U}_T$ governs the dynamics of $T_1$ and $T_2$. It determines their positions at each time step, following a movement pattern which can be deterministic or stochastic.

\subsubsection{Action Space $\mathcal{A}^i_{i \in I}$} 
At every time-step, agents decide in which direction along the $x$ and $y$ axis to move or if they should maintain their current position. We encode this action space in two dimensions, with each dimension having 3 options corresponding to going forward, backward, or hold along a certain axis.

\subsubsection{Transition Function \(\mathcal{P}\)}
The transition function \(\mathcal{P}\) defines the dynamics of the environment. It specifies the probability distribution over the next state \(s'\) given the current state \(s\) and the joint actions \(\mathbf{a}\) taken by the agents. 


In our scenario, the transition of states primarily depends on the agents' movements and the dynamic changes in the network topology due to the movement of the agents and targets. Specifically, when moving within the 2D plane, they alter their positions and potentially change the network's connectivity graph \(\mathcal{G}(t)\). These movements lead to a new set of edges \(\mathcal{E}(t+1)\) in the graph, to updated nodes' features, and, thus, to the transition to a new state \(s'\).



\subsubsection{Reward Function $\mathcal{R}$}
The reward function in our \gls{dpomdp} framework is designed to direct agents toward forming a communication network between moving targets. It consists of the following components:

\textbf{Base Connectivity Reward:}
The base reward, $R_{\text{base}}$, is given by the ratio of the number of nodes in the largest connected component and the total number of nodes:
\begin{equation}
    R_{\text{base}}(s) = \frac{|C_{\text{max}}(s)|}{|\mathcal{V}|},
\end{equation}
where $|C_{\text{max}}(s)|$ represents the size of the largest connected component in state $s$, and $|\mathcal{V}|$ is the total number of entities.


\textbf{Centroid Distance Penalty:}
The centroid distance penalty, $P_{\text{cent}}$, is computed as the Euclidean distance between the centroid of the agents' and targets' positions.

\textbf{Target Path Bonus:}
A bonus, $B_{\text{path}}=100$, is awarded if a path exists between the two targets.

\textbf{Overall Reward:}
The overall reward combines the base reward, centroid distance penalty, and the target path bonus:

\begin{equation}
    \label{eq:final_reward}
    R(s, a) = \begin{cases} 
    
    B_{\text{path}}(s) & \mbox{if } \exists \mathrm{path}(T_1, T_2);
    
    \\ R_{\text{base}}(s) - P_{\text{cent}}(s), & \mbox{otherwise}.
    \end{cases} 
\end{equation}

This reward ensures that agents are motivated to form a stable and efficient communication network while positioning themselves effectively relative to the moving targets.

\begin{table}[b]
    \centering
    \begin{tabularx}{\columnwidth}{ c | X | c | c | c | c}
\toprule
    Node Type & ID & Coord & Action & $T_1$ Coord & $T_2$ Coord \\
\midrule
    T & 4 & ($x_3$, $y_3$) & 0 & ($x_3$, $y_3$) & ($x_3$, $y_3$)\\
    A& 1 & ($x_1$, $y_1$) & 2 & ($x_3$, $y_3$) & ($x_3$, $y_3$)\\
    A & 2 & ($x_0$, $y_0$) & 7 & ($x_3$, $y_3$)& ($x_3$, $y_3$) \\
\bottomrule
\end{tabularx}
    \caption{Example of nodes' features in the graph representing some agent's neighborhood. The Action feature for Target nodes (T) is always set to 0.}
    \label{tab:state}
\end{table}


\begin{figure*}[t]
    \centering
    \begin{subfigure}[b]{0.6\textwidth}
        \includegraphics[width=0.9\textwidth]{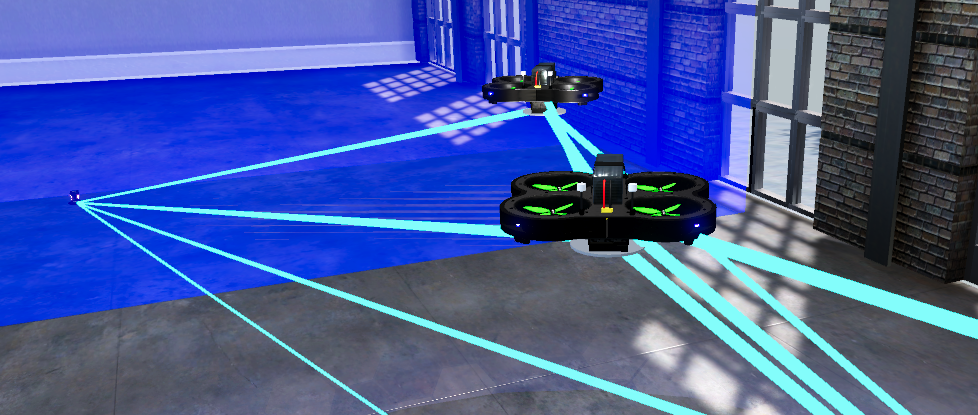}
        \caption{}
        \label{fig:figure3}
    \end{subfigure}
    \begin{subfigure}[b]{0.3\textwidth}
        \includegraphics[width=\textwidth]{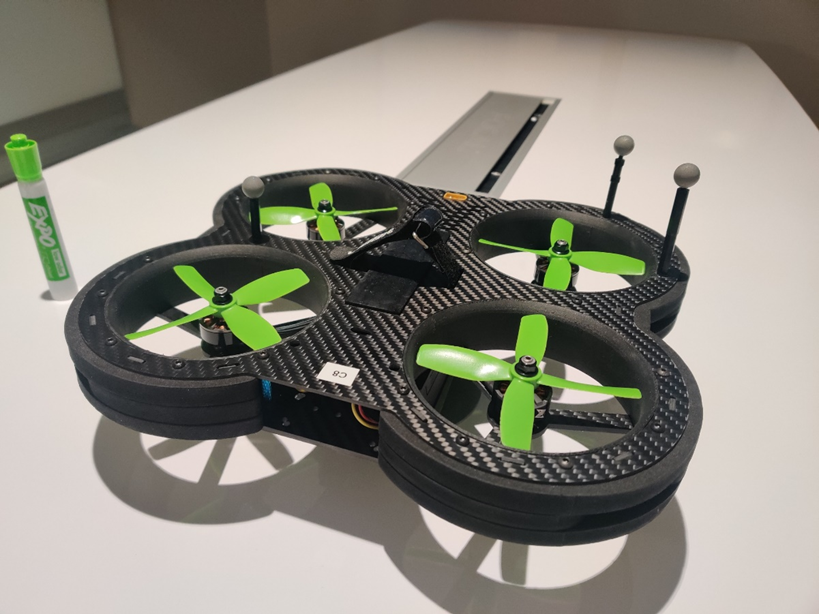}
        \caption{}
        \label{fig:figure1}
    \end{subfigure}
    \caption{A screen capture from our \gls{lvc} running our learned \gls{marl} strategies (a) and the corresponding real \gls{uav} (b).}
    \label{fig:three_figures}
\end{figure*}
\section{Learning Approach}
\label{sec:approach}
In this section, we describe our learning approach and the network architecture used to parameterize the agents' action-value function. Our approach leverages \gls{dgn}~\cite{Jiang2020Graph} in modeling relational data while harmoniously integrating with the decentralized and networked nature of the task. Training is performed in a \gls{ctde}~\cite{marl-book} fashion, with the agents optimizing the same action-value function parameterization.



\subsection{The Role of Message Passing and Latent Representations}
In \gls{dgn}, dot-product attention and graph convolution play a crucial role in integrating feature vectors associated with nodes within a local region around a certain node $i$, by generating a latent feature vector $h_i$ comprising node $i$ and its neighboring nodes. By adding more convolutional layers, the receptive field of an agent expands progressively, leading to the accumulation of more information. Consequently, the scope of cooperation also broadens, enabling agents to collaborate more effectively.
Specifically, with one convolutional layer, node $i$ aggregates the features of the nodes in its one-hop neighborhood.
When two layers are stacked, node $i$ receives the output of the first convolutional layer of nodes in its one-hop neighborhood, which, in turn, embeds information from nodes two hops away from $i$. However, irrespective of the number of convolutional layers, node $i$ only communicates with its one-hop neighbors, making \gls{dgn} practical in real-world networking scenarios.

\begin{figure}
    \centering
    \includegraphics[width=0.7\linewidth]{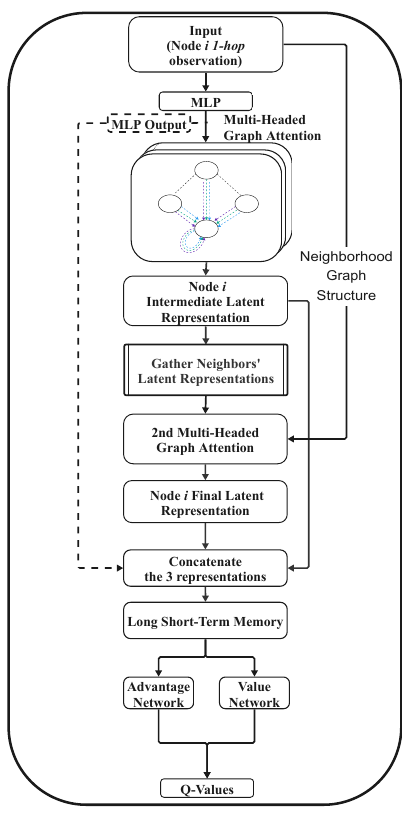}
    \caption{A Summary of our network architecture.}
    \label{fig:network_architecture}
\end{figure}

In this work, we achieve cooperation between the agents by enabling them to share their latent representations within their immediate neighborhood, conditioning their actions on such information. In addition, we also integrate target entities in the sharing process. Targets, despite not being learning agents, will also compute the latent representation of their one-hop neighborhood graph structure, in the same way agents do. To achieve this behavior, targets are deployed with the same \gls{gnn} architecture adopted by the agents, except for adopting only the first encoding module, which produces their intermediate latent representation (Figure \ref{fig:network_architecture}). Finally, if the neighborhood of a target entity $T_i$ includes the presence of one or more agents, these will be able to gather such representation and condition their actions accordingly.


\subsection{Integration of LSTM and Observation Stacking}
To handle the temporal dynamics and partial observability in our environment, we integrate \gls{lstm} networks employing graph observation stacking during training~\cite{HausknechtS15}. Observation stacking provides a richer representation of the environment by aggregating observations over multiple time steps, giving the \gls{lstm} the temporal context needed for effective learning. This combination allows our model to maintain a memory of past neighborhood observations, aiding in decision-making in a partially observable setting. 

\subsection{Summary of the Neural Network Architecture}
Our approach adopts the network architecture presented in Figure \ref{fig:network_architecture}, which comprises several components:
\begin{itemize}
    \item \textbf{Two Multi-Headed Graph Attention Layers:} Utilized for encoding the relational data among agents. These layers capture the spatial and relational dependencies in the network~\cite{veličković2018graph}. The first layer provides the latent representation to be shared with neighboring agents.
    \item \textbf{One \gls{lstm} Layer:} Used for capturing temporal dependencies and handling partial observability~\cite{HausknechtS15}.
    \item \textbf{Dueling Action Decoder:} Incorporates separate streams for estimating state values and advantages, to facilitate the estimation of the actions values~\cite{10.5555/3045390.3045601}.
\end{itemize}

Our model's architecture, with its \gls{gat} layer and \gls{lstm} integration, is tailored to address the challenges presented by dynamic network bridging with a cooperative perspective.

\section{Interaction with the Live Virtual Constructive UAV Framework}
\label{sec:lvc}
To enable a feasible and practical transition of learned \gls{marl} behaviors to real-world applications, we have been using a custom \gls{lvc} \gls{uav} environment developed in-house. The \gls{lvc} framework provides access to real and physically-accurate simulated \gls{uav} agents with agent-to-agent network communications. Figure~\ref{fig:three_figures} presents a screen capture of simulated \glspl{uav} and a picture of their real counterpart. By supporting simulated-only, real-only, or a mix of real and simulated \glspl{uav} operating in the same environment, our \gls{lvc} enables a multitude of opportunities for prototyping, evaluating, and deploying learned policies.

Each agent in the \gls{lvc} uses an onboard flight controller to maintain flight stability and an embedded Linux machine to deploy higher-level automation tasks such as the learned \gls{marl} behaviors. In the simulated environment, agents are deployed on individual Virtual Machines (VMs) that emulate the real embedded computer. Network communication between agents is achieved using the low-latency User Datagram Protocol (UDP) over a dedicated WiFi network. Network emulators such as the Extendable Mobile Ad-hoc Network Emulator (EMANE) are also supported, enabling emulation of realistic network conditions such as those encountered in disaster response situations. Once policies are trained within our \gls{marl} environment, we rigorously test them in the \gls{lvc} by interfacing the two components with gRPC and having the framework communicate the actions of each agent.
However, the \gls{lvc} also supports a deployment mode detached from our \gls{marl} framework, where policies are deployed in each real or simulated \glspl{uav} with agent-agent network communication enabled by the \gls{lvc}. Aircraft localization in real environments is achieved through a 12-camera Motion Capture (MoCap) system. Furthermore, the \gls{lvc} can adjust the localization accuracy and acquisition rates on the fly to simulate various sensors and conditions in a controlled environment.

In this work, the \gls{lvc} plays a fundamental role in the testing phase, assessing the policies' real-world performance, and evaluating their ability to navigate and interact effectively within dynamic real-world environments.




\begin{figure*}[t]
    \centering
    \begin{subfigure}[b]{0.3\textwidth}
        \includegraphics[width=\textwidth]{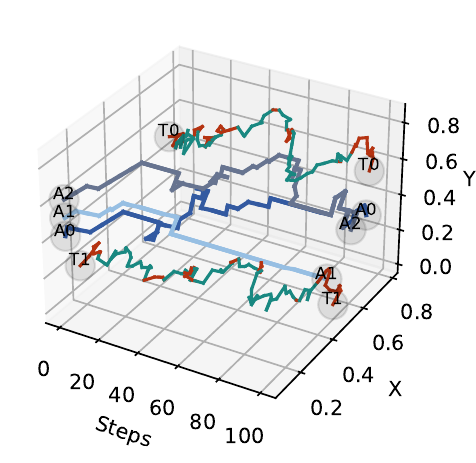}
        \label{fig:figure1}
    \end{subfigure}
    \hfill 
    \begin{subfigure}[b]{0.3\textwidth}
        \includegraphics[width=\textwidth]{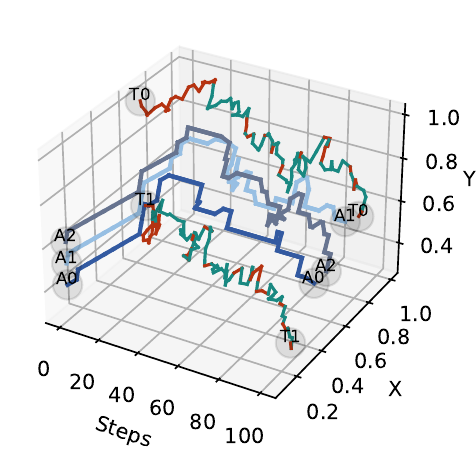}
        \label{fig:figure2}
    \end{subfigure}
    \hfill 
    \begin{subfigure}[b]{0.3\textwidth}
        \includegraphics[width=\textwidth]{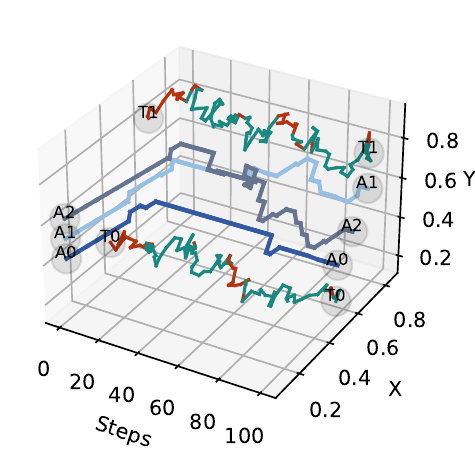}
        \label{fig:figure3}
    \end{subfigure}
    \caption{Examples of trajectories produced by agents and targets positions during 3 evaluation episodes. Green and Red segments (T1/T2) represent the intervals of (time)steps where the agents (A1, A2, A3) were able to form a link between T1 and T2. }
    \label{fig:results}
\end{figure*}

\section{Experiments}
\label{sec:experiments}
In this section, we will introduce our experimental setup, describing the steps involved in our training and evaluation process. Finally, we will present a comparison between our approach and a centralized heuristic, designed to reposition the agents according to the line connecting the two targets.

\begin{algorithm}
\caption{Centralized Heuristic (Baseline)}
\label{alg:heuristic}
\begin{algorithmic}[1]
\Require $\mathcal{I}, \mathcal{T}_1, \mathcal{T}_2, \mathcal{G}(t)$, step size for movement.
\Ensure Ideal Actions for each agent $i \in \mathcal{I}$.
\State $E \leftarrow $ [].
\State Determine the slope \( m \) and y-intercept \( b \) of the line connecting \( T_1 \) and \( T_2 \).
\State Evenly space \(|\mathcal{I}|\) x-values between \( x_1 \) and \( x_2 \).
\State Calculate the corresponding y-values using \( y = mx + b \) to form the set of potential endpoints \( P \).
\For{each agent $i \in \mathcal{I}$}
    \State Find the closest point $p \in P$.
    \State Add $p$ to $E$ and remove $p$ from $P$.
\EndFor

\For{each agent $i$, endpoint $e_i$ in $\mathrm{zip}(\mathcal{I}, E)$}
    \State Determine the relative position and orientation towards its assigned endpoint
    \State Calculate the directional angle $\alpha$ towards the endpoint.
    \If{$d(i, e_i) \le 0.1$}
        \State  Maintain current position.
    \Else
        \State Assign discretized action based on $\alpha$.
    \EndIf
\EndFor
\end{algorithmic}
\end{algorithm}

\subsection{Experimental Setup}

In our experiments, three agents and two targets are initially positioned on a normalized map with axes ranging from 0 to 1 to simulate the operational environment for the agents and targets. Each agent and target within this environment has a communication range set to 0.25, ensuring a functional scope for link formation. The step size for both agents and targets was set to 0.05, allowing them to move incrementally toward a chosen direction. However, the next position entailed by the agents' action can be interpreted as the next waypoint to which the agent will be directed. Such consideration allows our approach to abstract various elements, such as the agents' speed and route to the chosen waypoint, which can be determined by other policies. In this work, we leveraged internal mechanisms implemented in the \gls{lvc}.

Our training regime involved 33000 episodes across randomly generated scenarios, with every episode involving 100 decisions by each of the three agents. During both the training and testing phases, agents were consistently deployed at predetermined positions, resembling their base station. Initial positions were located at (0.1, 0.42), (0.1, 0.52), and (0.1, 0.62) for agents 1, 2, and 3 respectively.

During training, two targets were placed randomly on the map, with the constraint that the distance between them ranged between 0.5 and 0.7. This constraint was enforced to avoid overly simplistic or impossible scenarios, ensuring a balanced difficulty level for the agents to form a communication link. The movement of these targets was governed by a seeded random number generator, which was instrumental in selecting their next move. This random yet controlled movement ensured that the targets maintained the required distance apart, presenting the agents with realistic and challenging scenarios.

The learned multi-agent strategies were evaluated across 100 generated scenarios. These scenarios were controlled with seeded randomness for the placement and movement of the targets, ensuring the reproducibility of the experiments. Additionally, for evaluation purposes, our methods were compared against an ideal centralized heuristic, designed to coordinate the agents optimally (Algorithm \ref{alg:heuristic}).


\begin{figure}[t]
    \centering
    \includegraphics[width=0.9\linewidth]{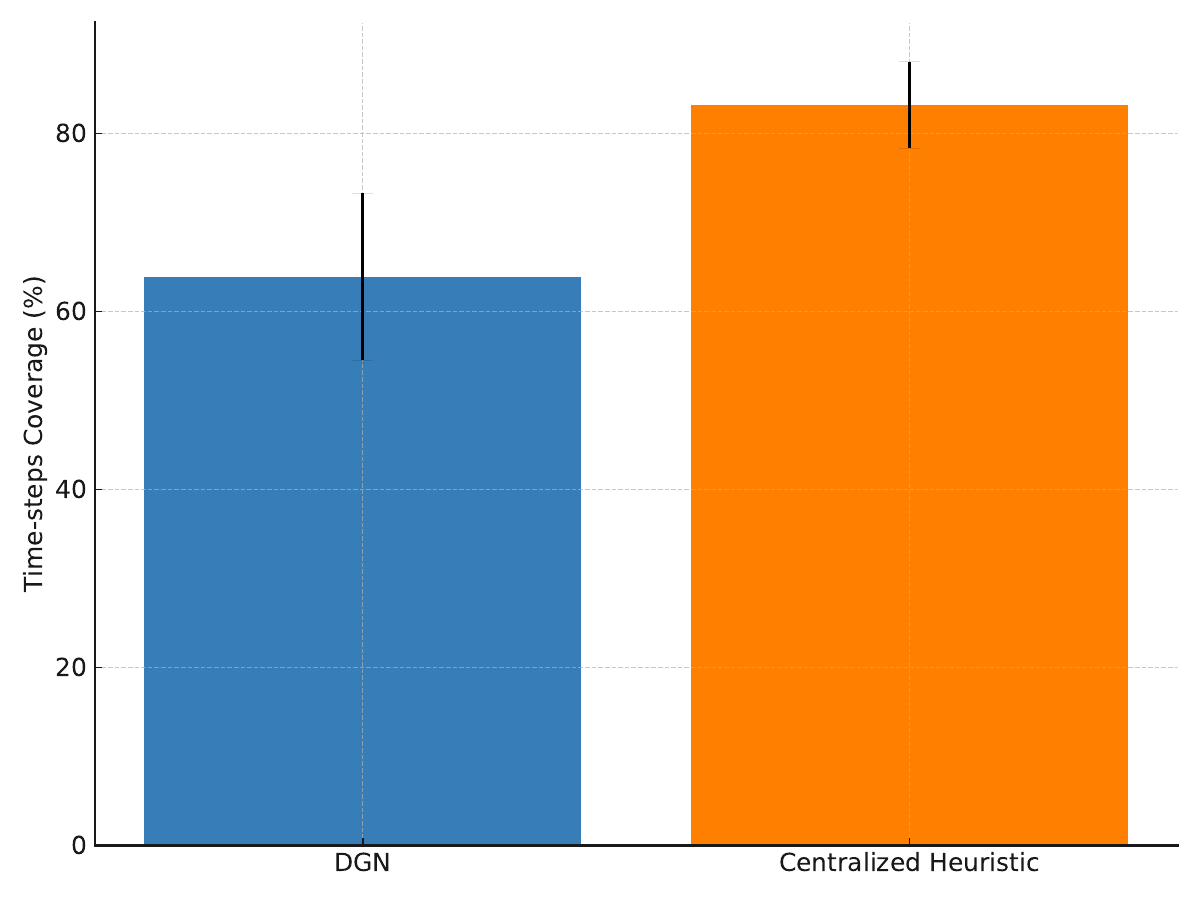}
    \caption{Comparison in terms of average time-steps covered during the evaluation phase of our agent and the centralized heuristic.}
    \label{fig:comparison}
\end{figure}
\subsection{Results}

Figure \ref{fig:results} illustrates the trajectories of agents and targets throughout three different episodes, each lasting 100 time-steps. The learned strategies enable the agents to form a communication link between $T_1$ and $T_2$ for most of the episode duration, despite the unpredictable movements of the targets. Figure \ref{fig:comparison} offers a performance comparison between our \gls{dgn} approach and the centralized heuristic in terms of "time-steps coverage," which is defined as the percentage of time-steps during which a path between $T_1$ and $T_2$ is successfully maintained. The \gls{dgn} agents managed to bridge the two targets for $63.88\%$ of the episode duration, while the optimal heuristic achieved this task for $83.19\%$ of the time-steps. Additionally, the average total return was $6494.13 \pm 941.50$ for the \gls{dgn} method and $8440.28 \pm 490.58$ for the heuristic. We note that given the limited step size and the initial starting position of the agents, significantly distant from the two targets, it is not possible to achieve network bridging during the entire episode. 
Despite the performance of our agents being lower than the one achieved by the centralized heuristic, our initial results underscore the potential of our \gls{dgn} approach in dynamic and unpredictable environments, opening avenues for further optimization and application in real-world networks.

\section{Future Work}
    To enrich the practical significance of our \gls{marl} approach, we plan to incorporate low-level networking elements in our learning environment. To this end, we will host our learning agents in container-based network simulators/emulators~\cite{10356270} and include ray tracing propagation models to better mirror realistic scenarios~\cite{10356208}. This will allow us to directly optimize networking properties related to the connectivity of the underlying network. We will also enable our agents to take more refined actions by including continuous action spaces and investigating learning frameworks based on entropy-regularized stochastic policies. The first will allow the agents to choose any point within the circle/sphere bounded by the step size as their next waypoint, while the second will encourage the learning of multi-modal strategies. Additionally, integrating our learning framework with established control systems~\cite{10383589} will be explored to support safety and reliability for real-world applications, ensuring our advancements are both innovative and grounded for practical deployment. 

Finally, we conjecture that learned multi-agent strategies will increase their benefit when introduced to more complex network bridging scenarios presenting numerous target entities. In these settings, employing centralized/distributed heuristic functions will present more challenges and limitations than in instances where only two target nodes are involved.



\section{Conclusion}
In this work, we proposed a \gls{dpomdp} formulation for dynamic network bridging and we showed how effective decentralized multi-agent strategies involving proactive communication can be learned for this task. Furthermore, we have designed a \gls{marl} approach to solve the problem in a distributed manner and we compared our results to a more traditional centralized approach. The experimental results showed promising results in this direction, with our agents being able to form a link between the targets in every training episode, for the most time of the episodes' duration.

\bibliographystyle{IEEEtran}
\bibliography{bibliography}

\begin{thebibliography}{10}
\providecommand{\url}[1]{#1}
\csname url@samestyle\endcsname
\providecommand{\newblock}{\relax}
\providecommand{\bibinfo}[2]{#2}
\providecommand{\BIBentrySTDinterwordspacing}{\spaceskip=0pt\relax}
\providecommand{\BIBentryALTinterwordstretchfactor}{4}
\providecommand{\BIBentryALTinterwordspacing}{\spaceskip=\fontdimen2\font plus
\BIBentryALTinterwordstretchfactor\fontdimen3\font minus \fontdimen4\font\relax}
\providecommand{\BIBforeignlanguage}[2]{{%
\expandafter\ifx\csname l@#1\endcsname\relax
\typeout{** WARNING: IEEEtran.bst: No hyphenation pattern has been}%
\typeout{** loaded for the language `#1'. Using the pattern for}%
\typeout{** the default language instead.}%
\else
\language=\csname l@#1\endcsname
\fi
#2}}
\providecommand{\BIBdecl}{\relax}
\BIBdecl

\bibitem{Jiang2020Graph}
\BIBentryALTinterwordspacing
J.~Jiang, C.~Dun, T.~Huang, and Z.~Lu, ``Graph convolutional reinforcement learning,'' in \emph{International Conference on Learning Representations}, 2020. [Online]. Available: \url{https://openreview.net/forum?id=HkxdQkSYDB}
\BIBentrySTDinterwordspacing

\bibitem{veličković2018graph}
\BIBentryALTinterwordspacing
P.~Veličković, G.~Cucurull, A.~Casanova, A.~Romero, P.~Liò, and Y.~Bengio, ``Graph attention networks,'' in \emph{International Conference on Learning Representations}, 2018. [Online]. Available: \url{https://openreview.net/forum?id=rJXMpikCZ}
\BIBentrySTDinterwordspacing

\bibitem{HausknechtS15}
M.~J. Hausknecht and P.~Stone, ``Deep recurrent q-learning for partially observable mdps,'' in \emph{2015 {AAAI} Fall Symposia, Arlington, Virginia, USA, November 12-14, 2015}.\hskip 1em plus 0.5em minus 0.4em\relax {AAAI} Press, 2015, pp. 29--37.

\bibitem{Zhou2015}
\BIBentryALTinterwordspacing
Y.~Zhou, N.~Cheng, N.~Lu, and X.~S. Shen, ``Multi-uav-aided networks: Aerial-ground cooperative vehicular networking architecture,'' \emph{IEEE Vehicular Technology Magazine}, vol.~10, pp. 36--44, 12 2015. [Online]. Available: \url{http://ieeexplore.ieee.org/document/7317860/}
\BIBentrySTDinterwordspacing

\bibitem{Tortonesi2012}
M.~Tortonesi, C.~Stefanelli, E.~Benvegnu, K.~Ford, N.~Suri, and M.~Linderman, ``Multiple-uav coordination and communications in tactical edge networks,'' \emph{IEEE Communications Magazine}, vol.~50, pp. 48--55, 10 2012.

\bibitem{Park2016DroneFA}
S.~J. Park, H.~Kim, K.~Kim, and H.~Kim, ``Drone formation algorithm on 3d space for a drone-based network infrastructure,'' \emph{2016 IEEE 27th Annual International Symposium on Personal, Indoor, and Mobile Radio Communications (PIMRC)}, pp. 1--6, 2016.

\bibitem{Hammami2019}
S.~E. Hammami, H.~Afifi, H.~Moungla, and A.~Kamel, ``Drone-assisted cellular networks: A multi-agent reinforcement learning approach,'' \emph{IEEE International Conference on Communications}, vol. 2019-May, 5 2019.

\bibitem{Jiang2021}
B.~Jiang, S.~N. Givigi, and J.~A. Delamer, ``A marl approach for optimizing positions of vanet aerial base-stations on a sparse highway,'' \emph{IEEE Access}, vol.~9, pp. 133\,989--134\,004, 2021.

\bibitem{marl-book}
\BIBentryALTinterwordspacing
S.~V. Albrecht, F.~Christianos, and L.~Sch\"afer, \emph{Multi-Agent Reinforcement Learning: Foundations and Modern Approaches}.\hskip 1em plus 0.5em minus 0.4em\relax MIT Press, 2023. [Online]. Available: \url{https://www.marl-book.com}
\BIBentrySTDinterwordspacing

\bibitem{gnn}
F.~Scarselli, M.~Gori, A.~C. Tsoi, M.~Hagenbuchner, and G.~Monfardini, ``The graph neural network model,'' \emph{IEEE Transactions on Neural Networks}, vol.~20, no.~1, pp. 61--80, 2009.

\bibitem{Busoniu2010}
L.~Bu{\c{s}}oniu, R.~Babu{\v{s}}ka, and B.~De~Schutter, \emph{Multi-agent Reinforcement Learning: An Overview}.\hskip 1em plus 0.5em minus 0.4em\relax Berlin, Heidelberg: Springer Berlin Heidelberg, 2010, pp. 183--221.

\bibitem{10.5555/3045390.3045601}
Z.~Wang, T.~Schaul, M.~Hessel, H.~Van~Hasselt, M.~Lanctot, and N.~De~Freitas, ``Dueling network architectures for deep reinforcement learning,'' in \emph{Proceedings of the 33rd International Conference on International Conference on Machine Learning - Volume 48}, ser. ICML'16.\hskip 1em plus 0.5em minus 0.4em\relax JMLR.org, 2016, p. 1995–2003.

\bibitem{10356270}
R.~Galliera, M.~Zaccarini, A.~Morelli, R.~Fronteddu, F.~Poltronieri, N.~Suri, and M.~Tortonesi, ``Learning to sail dynamic networks: The marlin reinforcement learning framework for congestion control in tactical environments,'' in \emph{MILCOM 2023 - 2023 IEEE Military Communications Conference (MILCOM)}, 2023, pp. 424--429.

\bibitem{10356208}
A.~Amato, R.~Fronteddu, and N.~Suri, ``Dynamically creating tactical network emulation scenarios using unity and emane,'' in \emph{MILCOM 2023 - 2023 IEEE Military Communications Conference (MILCOM)}, 2023, pp. 201--206.

\bibitem{10383589}
A.~Chen, K.~Mitsopoulos, and R.~Romagnoli, ``Reinforcement learning-based optimal control and software rejuvenation for safe and efficient uav navigation,'' in \emph{2023 62nd IEEE Conference on Decision and Control (CDC)}, 2023, pp. 7527--7532.

\end{thebibliography}

\end{document}